\documentclass[final]{aipproc}

\layoutstyle{6x9}

\begin{document}

\title{Radiation-induced solitary waves in hot plasmas of accretion disks}

\classification{52.25.Qs, 52.35.Sb, 95.30.Gv, 95.30.Jx.}

\keywords{accretion, accretion disks}

\author{Fedor V.Prigara}{address={Institute of Microelectronics and Informatics,
Russian Academy of Sciences,\\ 21 Universitetskaya, 150007
Yaroslavl, Russia}}

\begin{abstract}

It is shown that the existence of radiation-induced solitary waves
in hot plasmas of accretion disks depends on the radial
temperature profile.

\end{abstract}

\maketitle

A hot plasma is four-component medium which includes the electron
and ion fluids, magnetic field, and radiation. Each of these
components produces its own branch of linear and nonlinear waves
in plasmas. The nonlinear branches include in particular
radiation-induced density waves in hot plasmas which are
considered below.

Recently, the energy distribution of atoms in the field of thermal
black body radiation was obtained \cite{Prigara03} in the form

\begin{equation}
\label{eq1}
N/N_{0} = \sigma _{a} \omega ^{2}/\left( {2\pi c^{2}} \right)\left(
{exp\left( {\hbar \omega /kT} \right) - 1} \right),
\end{equation}

\noindent where $N_{0} $ is the population of the ground state
$E_{0} $, $N $ is the population of the energy level $E = E_{0} +
\hbar \omega $, $\sigma _{a} $ is the absorption cross-section,
$\hbar $ is the Planck constant, and $T $ is the radiation
temperature.

The function (1) has a maximum at $\hbar \omega _{m} = 1.6kT$.
When the temperature exceeds the critical value of $T_{0} = 2
\times 10^{7^{}}K$ (the inversion temperature), the population of
the energy level $E $ exceeds the population of the ground state
$E_{0} $. Since the function (1) is increasing in the range
$\omega < \omega _{m} $ , the inversion of the energy level
population is produced also in some vicinity of $\omega _{m} $
(below $\omega _{m} $). This suggests the maser amplification of
thermal radio emission in continuum by a hot plasma with the
temperature exceeding the critical value $T_{0} $. Since a hot
plasma in an accretion disk is concentrated nearby the central
energy source, maser amplification is characteristic for compact
radio sources.

It was shown recently \cite{Prigara03} that thermal emission from
non-uniform gas is produced by an ensemble of individual emitters.
Each of these emitters is an elementary resonator the size of
which has an order of magnitude of mean free path $l $ of photons,

\begin{equation}
\label{eq2} l = 1/n\sigma,
\end{equation}

\noindent where $n $ is the number density of particles and
$\sigma $ is the absorption cross-section. The absorption
cross-section for ions is given by the formula $\sigma _{i} = \pi
\left( {8\pi \hbar ^{3}c/me^{4}} \right)^{2}$, where \textit{m} is
the mass and \textit{e} is the charge of an electron respectively,
and \textit{c} is the speed of light. The absorption cross-section
for neutral atoms is $\sigma _{0} \cong 10^{ - 15}cm^{2}$.

The emission of each elementary resonator is coherent, with the
wavelength $\lambda = l $, and thermal emission of gaseous layer
is incoherent sum of radiation produced by individual emitters.

If the temperature of plasma exceeds the inversion temperature,
$T_{0} $, then the thermal radiation from an elementary resonator
is amplified by the laser mechanism and emitted in the direction
of decreasing gas density.

The time of life for an elementary resonator has an order of magnitude

\begin{equation}
\label{eq3}
\tau \cong l/v_{iT} \cong l/c_{s} ,
\end{equation}

\noindent where $v_{iT} $ is the ion thermal velocity, and $c_{s}
$ is the speed of sound. Such is the time duration of the light
pulse produced by an elementary resonator.

The absorption of this radiation pulse by the nearby gaseous layer
leads to the heating of a plasma and its density perturbations.
The radiation pulse is then re-emitted by the new virtual
elementary resonator in a plasma. In such a manner, the
radiation-induced density wave propagates through the hot plasma.

Various astrophysical objects, such as active galactic nuclei,
X-ray binaries, young pulsars, have hot accretion disks. Since the
density gradient in an accretion disk is directed along the
radius, the radiation-induced density wave in a hot plasma of an
accretion disk normally propagates as the radial density wave.

The inversion of energy level population associated with the
radial density wave produces a moving pulse of coherent radiation
with the changing wavelength determined by equation (2). Such are
the radio pulses from pulsars \cite{Prigara04}.

In fact, the inversion of energy level population corresponding to
the radio band (in some vicinity of $\omega _{m} $ below it)
exists also at the temperatures below the inversion temperature,
$T_{0} $. This conclusion is confirmed by the temperature profile
in active galactic nuclei which is virial, i.e.

\begin{equation}
\label{eq4}
T \propto r^{ - 1},
\end{equation}

\noindent where $r $ is the distance from the central energy
source.

However, there are no observed pulsed emission from the active
galactic nuclei, contrary to the pulsars. Since the temperature
profile in pulsars is different from those given by equation (4)
and has a form \cite{Prigara04}

\begin{equation}
\label{eq5}
dT/dr > 0,
\end{equation}

\noindent it is clear, that a hot plasma is required for
triggering of radio and X-ray pulses.

\end{document}